# Affordance switching in self-organizing brain-body-environment systems

Vicente Raja[1,2] & Matthieu M. de Wit[3]

[1]Department of Philosophy, University of Murcia, Murcia, Spain; [2]Rotman Institute of Philosophy, Western University, London, ON, Canada; [3]Department of Neuroscience, Muhlenberg College, Allentown, PA, USA

Correspondence: vicente.raja@um.es (V. Raja); matthieudewit@muhlenberg.edu (M. M. de Wit).

ORCIDs: 0000-0001-5780-9052 (V. Raja); 0000-0002-9964-1339 (M. M. de Wit).

## Abstract

In the ecological approach to perception and action, information that specifies affordances is available in the energy arrays surrounding organisms, and this information is detected by organisms in order to perceptually guide their actions. At the behavioral scale, organisms responding to affordances are understood as self-organizing and reorganizing softly-assembled synergies. Within the ecological community, little effort has so far been devoted to studying this process at the neural scale, though interest in the topic is growing under the header of ecological neuroscience. From this perspective, switches between affordances may be conceptualized as transitions within brain-body-environment systems as a whole rather than under the control of a privileged (neural) scale. We discuss extant empirical research at the behavioral scale in support of this view as well as ongoing and planned work at the neural scale that attempts to further flesh out this view by characterizing the neural dynamics that are associated with these transitions while participants dynamically respond to affordances.

**Glossary**

*Note that several of the terms in this glossary feature in the definitions of other terms. Hence, it may be helpful to look at those definitions together.*

**Behavioral scale:** The scale of description and/or analysis of the complete behaving organism in its environment. This can be contrasted with describing or analyzing a scale nested within this larger scale such as the neural scale. The behavioral scale itself can also be nested in yet larger scales such as, e.g., cultural settings in the case of human beings.

**Brain-body-environment system***:* Similar to organism-environment system. Brain and body are listed separately in this case to emphasize that each can be an object of study—and should be, in a complete science of behavior. A more appropriate description of "body" in this nomenclature would be "non-neural body" given that brain (and spinal cord and peripheral nervous system, for that matter) are of course also part of the body. The hyphens help emphasize that the contributions to behavior of each component cannot be understood outside of the context of the contributions of each of the other components, and that the components are tightly integrated. In fact, it is possible for synergies within brain-body-environment systems to crosscut components such as during competent tool use, showing that the boundary that is implied by listing the components separately is in some sense artificial.

**Ecological neuroscience:** An approach to neuroscience that integrates the study of the nervous system into the ecological approach to perception and action. Neuroscientific research can be classified as ecological when it treats the brain-body-environment system, rather than just the brain, as the unit of analysis. Another way of saying this is that ecological neuroscience studies the nervous system not in isolation but in the context of embodied, embedded organismal activity.

**(Enabling) constraints:** Constraints are usually regarded as limitations that impede one or another capacity of a given system. These limitations can be physical, biological, social, legal, and the like. Enabling constraints are limitations that result in functional capacities of a system. For instance, the anatomical limitations of human knees enable tractable motor control in walking and bipedalism. In this case, an anatomical *constraint enables* a function.

**Intrinsic mechanistic explanations:** Explanations of behavior that point exclusively to processes internal to the organism, typically neural processes, thereby ignoring contributions of the non-neural body and environment.

**(Neural and non-neural) resources:** Neural resources for a particular behavior refer to the neural components involved in the behavior. For instance, the activity of the hippocampus involved in memory and remembering. Non-neural resources refer to those non-neural components involved in the behavior. For instance, the agenda of the cell phone involved in memory and remembering the phone number of our contacts.

Resources typically have functional dispositions, as opposed to functional specificities, meaning that they are more likely to participate in some behaviors than others, but are not necessarily specialized *for* those behaviors.

**Organism-environment system:** A system that includes both the organism and its environment. When a phenomenon is a property of the organism-environment system, this means that it can only be studied and understood by simultaneously referring to both the organism and its environment. For example, to understand whether an affordance is present for an organism, one needs to characterize both the environmental context and the action capabilities of that particular organism.

**Phenotypic reorganization:** A name for the way in which biological organisms organize themselves for various behaviors: walking, running, eating, grasping, throwing, etc.

**Privileged (or dominant) scale:** A scale is privileged if it organizes activity in each of the other scales within a system. As an example, within mainstream cognitive neuroscience, the neural scale is viewed as privileged in that it is thought to control the bodily activity observed at the behavioral scale.

**Redundancy:** A system is redundant if it is able to express the same behavior in more than one way, involving different resources. Redundancy is closely related to degeneracy. However, the two concepts are distinct in that redundancy refers to the presence of an alternative behavioral solution that involves structurally similar components (e.g., different sets of neurons, muscle fibers, or other elements of the same type), while degeneracy refers to a situation in which an alternative behavioral solution involves (at least in part) structurally dissimilar components (e.g., locomotion via walking or pogo stick). Alternative solutions that involve the same components organized in different ways are also often referred to as redundant (e.g. locomotion via walking or crawling both involve the legs though they make a different contribution in each case). The questions of what constitutes structural similarity and dissimilarity and what constitutes a component are areas of active philosophical research.

**Self-organization:** Organization of a system that is not explained by descriptions of activity at a privileged (e.g., neural) scale internal to the system, nor by an external coordinator. Self-organized behavior emerges from interactions between components of the system at all scales.

**Soft-assembly:** A system is softly-assembled when it expresses a behavior via temporary, flexible, typically redundant, self-organized resources in the context of task-relevant (enabling) constraints. This can be contrasted with a hard-assembled system, which is composed of functionally-specialized elements that are typically organized by a privileged scale.

**Synergy:** A group of elements that work together as a functional unit to realize a given behavior.

In this chapter, we develop answers to the five questions posed to each of the contributors of this volume on the modern legacy of James Gibson's affordance concept. We devote special attention to the important but relatively unexplored question of how organisms switch from responding from one to another affordance, focusing in particular on understanding the dynamics at the neural scale within the brain-body-environment system during this process.

**What do we understand by the term *affordance*?**

We understand affordances as opportunities for action that organisms find in their environment. These opportunities for action are things like the *grab-ability* of a mug, the *pass-ability* of a door, the *catch-ability* of a ball, the *cross-ability* of a street or a river, etc. The presence of an affordance for an organism in a given situation can be arbitrarily complex. For instance, the grab-ability of a mug depends on the size and shape of the mug but also on the anatomy and skills of the organism. And the case of the cross-ability of a street is even more complicated as, at least for humans, it involves social norms and constraints on top of the environmental layout and the body and abilities of the street crosser. While acknowledging this complexity, ecological psychologists claim organisms can directly perceive affordances because there's enough environmental information as to that to be the case (see the next Section for brief further discussion). Direct perception of affordances and the environmental availability of information to do so are the only two constraints of our understanding of the concept.

This means we will not engage in definitional debates regarding the ontological status of affordances. There are several options out there: affordances as dispositions (Turvey et al., 1981; Turvey, 1992), affordances as nonfactual dispositions (Heras-Escribano, 2019), affordances as relations (Chemero, 2003; Stoffregen, 2003), and affordances as their own thing (Barrett, 2020), to name a few. If you push either of us, we'd likely choose one of these ontological characterizations of affordances for one or another reason. However, we think this is not relevant for our current purposes. The analysis we provide in the following sections should be immune to these ontological considerations. All the different ontological positions should be compatible with the idea of affordances as directly perceivable (via information) opportunities for action.

**What role does affordance play in perception? Is it the entity organisms perceive, or is it the means through which organisms perceive? Are affordances the only perceptual dependent variables?**

Affordances are objects of perception. Other theories of perception, including the classical ones, take colors, distances, or shapes, for instance, to be the objects of perception. The novelty introduced by James Gibson in ecological psychology was to include opportunities for action in the list of the possible objects of perception–along with removing some of the classical ones from that same list. In this context, affordances are not the means through which organisms perceive. If anything, the means through which organisms perceive affordances in their environment have to do with the detection of information available in the energy arrays (e.g., light, chemicals) that surround them. The way such information specifies the available affordances and the way organisms are able to detect it are open empirical questions on which ecological psychologists keep working (see, e.g., Lee, 2009; Raja, 2020; Segundo-Ortín, Heras-Escribano, & Raja, 2019; Turvey, 2018). But that there is (ecological) information that specifies affordances and that such information is detectable by organisms is perhaps the core ecological hypothesis:

> … [I]f there's information in light for the perception of surfaces, is there information for the perception of what they afford? Perhaps the composition and layout of surfaces *constitute* what they afford. If so, to perceive them is to perceive what they afford. This is a radical hypothesis, for it implies that the "values" and "meanings" of things in the environment can be directly perceived. (Gibson, 1979/2014, p. 119; emphasis in the original).

There are two dimensions to this quote. First, it restates the hypothesis of affordances as objects of perception. The information in light–Gibson was focused on vision–is, perhaps, for affordances. This is a way to say ecological information specifies affordances. And second, it seems to clearly suggest affordances are not the *only* object of perception. Surfaces are perceivable as well, or at least that's also explicitly claimed. And events are generally understood as objects of perception as well in the ecological literature (see Shaw & Cutting, 1980; Stoffregen, 2000). Also, places themselves have been proposed as perceivable (Heft, 2018).

An open question is also suggested in the quote above: are surfaces and their affordances the *same*? If that were the case, affordances could be in some sense the *only* object of perception. We would only need to equate not only surfaces, but also events and places, with affordances and we would have just one perceptual object to bound them all. To the best of our knowledge, however, this kind of thesis has not been properly defended in the literature. So, pending a convincing argument that defends otherwise, we hold a pluralist position regarding the objects of perception.

**Why is affordance any better than stimulus? What does a theory of affordance suggest that stimuli cannot; how has it moved the needle past Gibson 1960's recognition of how little we can define stimuli?**

We want to begin this section with a cautionary note: comparing the notions of affordance and stimulus can be misleading. The role affordances fulfill in the ecological theory is not completely akin to the role stimulus fulfill in other theories. As noted above, affordances are usually understood as one of the objects of perception; that is, one of those aspects of the environment organisms perceive by detecting ecological information. The way stimuli are characterized in the cognitive sciences writ large is more complicated than this. As James Gibson (1960) pointed out, the different uses of the notion of stimulus in the literature are not always compatible and, more generally, do not offer a coherent understanding of what stimulation *is* and *how* it is related to perception and action. Raja (2022) reviews a bit of contemporary literature in which very different entities such as the input of retinal ganglion cells (Sayood, 2018), the vibration of rats' whiskers (Ince et al., 2010), the structural properties of pictures of faces (Hashemi, Pachai, Bennett, & Sekuler, 2019) or the movie *Memento* (Kauttonen, Hlushchuk, Jääskeläinen, & Tikkaac, 2018) are labeled as "stimulus". Some of these entities seem to refer to whatever triggers the activation of a physiological receptor (e.g., the input of retinal ganglion cells) while some others seem to refer to whatever our perceptual experience–psychological experience, more generally–of a given sensory input is (e.g., the movie *Memento*). This variety of conceptions of stimulation reinforces Gibson's claims in the 1960s and shows that the last few decades have not witnessed a serious improvement on the issue. Moreover, it highlights the mismatch between the notions of affordance and stimulus.

The disconnection between affordances and stimuli is clear when the latter are understood in physiological terms. What triggers the activation of a physiological receptor is *not* the object of perception. In the mainstream theory, whatever inputs our sensory receptors get are not enough to provide perceptual knowledge of the environment, and perceptual states are only possible after the sensory input is processed by the brain. Perceptual states are built up within the brain in the form of perceptual models or perceptual representations (see the next Section for further discussion of explanation via intrinsic mechanisms). And the objects of perception that feature in those representations are not stimuli, but things, distances, scenes, etc. The physiological notion of stimuli has been historically applied *tout court* into psychology and has resulted in the idea of *atomic* stimulation. Put simply, atomic stimulation in psychology is physiological stimulation with sensory correlates–e.g., correlates are not activations of receptors but simple sensations such as "cold" or "yellow" or "C sharp".

Gibson followed a tradition that can be traced back to William James and his defense of the richness and complexity of stimulation. Stimulation can be rich (i.e., non-atomic) and it can actually provide perceptual knowledge about the environment without the need for internal processing or combination of simple sensations. The ecological notion of stimulus information is a culmination of this kind of thinking. So the story goes, some structures of the arrays of stimulation that surround us and other organisms are informative of the environment and our (and their) relation with it. For instance, together with other sources of information (e.g., inertial), a global centrifugal flow in our optical field tells us we are engaging in forward locomotion, while a global centripetal flow tells us we are locomoting backwards. These (higher-order) structures are *ecological information.* For ecological psychologists, ecological information provides organisms with perceptual contact to their environment by specifying the opportunities for interaction the latter offers to them. These opportunities for interaction are, of course, affordances. Ecological information, as a form of stimulus information, substitutes the classical notion of atomic stimuli in the context of ecological psychology and provides a naturalistic way to account for perception without recourse to neural processes of construction, combination, or enrichment.

But what about the other psychological notion of stimulus? Aren't affordances like the objects of perception we experience when we watch *Memento*, for instance? In a way, affordances are just a *different* kind of object. Therefore, if *Memento* is a stimulus, affordances must somehow be a stimulus as well. According to Gibson (1960), this conflates the notion of stimulus with the notion of object of perception or object of experience. It is just a different uncareful application of the physiological notion of stimulus in psychology, this time not focusing on its atomistic nature but on its dependence on the response triggered–e.g., the movie *Memento* is a stimulus because the experience of watching *Memento* is triggered by it. Despite its problems, and even if we do not endorse it, this is the accepted status quo in the field. In a strict sense, affordances as objects of perception can be regarded as "stimuli" as much as any other object of perception. Affordances, nevertheless, are still different from those other objects of perception that mainstream cognitive sciences label as stimuli: while all the mainstream objects of perception achieve their perceptual status by a process of mental construction or enrichment, affordances are objects of perception because they are specified by ecological information. This, again, is the core thesis of ecological realism. Affordances are properties of the organism-environment system that are meaningful to the organism without the need for a subjective attribution of meaning. Affordances provide a theory of meaning that takes meaning from "inside here", that is, from the minds (or brains) of perceivers, and puts it "out there", that is, in the organism-environment relationship. This is a revolution in psychology that would be impossible without abandoning the classical notion of stimulus.

**What is the connection between affordances and behavioral scale and intention? How distant or reluctant can a perceiving-acting system be before affordances come into play? Do affordances exert their role only when the system is close or willing enough?**

When organisms act in their environment, they perceptually guide their own action with respect to affordances. In this context, a complete description of the behavioral scale includes the affordances involved in the described behavior. At least according to ecological psychologists, affordances are therefore tightly connected to behavior and to the behavioral scale of analysis (cf. de Wit, de Vries, van der Kamp, & Withagen, 2017).

The relationship between affordances and intentions is not as straightforward as the one between affordances and behavior. The most important reason for this is perhaps that intentions are difficult to operationalize. Simply defined, intentions are the goals or aims of a cognitive system. They are a name for all those psychological states related to the willingness *to do* something. In a sense, intentions are triggers of behavior: we walk, run, grab a mug, kick a ball, write a poem, and so on, because we intend to do so. It is relatively easy to agree with such a simple definition when in a colloquial environment. However, it is not that easy to make scientific sense of it. This difficulty is likely due to the dual character of intentions that is revealed after a deeper scrutiny of the concept. On the one hand, intentions seem to be *intrinsic* to the cognitive system. Cognitive systems seem to be the origin of their own intentions and, therefore, of their own behavior. On the other hand, intentions seem to *transcend* cognitive systems. At least sometimes, intentions are about things other than cognitive systems–such aboutness is usually called "intentionality" by philosophers–and we recognize them as produced by extrinsic factors. In this sense, intentions seem to belong to the cognitive system and, at the same time, they seem to point to other things outside that system. Therefore, the main question is: how can we make sense of this duality scientifically?

It is not possible to provide a general answer to this question in the rest of this chapter. A possible way to narrow its scope into a manageable project is to address intentions from the point of view of task switching. How do we change from performing one task to performing another? Imagine you take the tram and decide to remain standing up despite the availability of free seats. After a couple of stops, you decide to sit down on one of those free seats. What is going on for that change to occur? Or imagine that, while playing a football game, you are running with the ball and then pass the ball to another member of your team. What happened for you to change from running to passing at that very moment? What is going on in these two situations for you to switch from one task (standing; running) to a different one (sitting; passing)? It seems that, maybe along with other cognitive and emotional processes, these are events in which cognitive systems change or develop their intentions. In this context, one possibility is to describe a fully intrinsic mechanism to explain task switching. Such a mechanism would entail the complete intrinsicality of intentions. This is a common strategy in cognitive neuroscience. For instance, Vinod Menon and Lucina Uddin (2010) have proposed the insula as a multitasking hub in the brain in charge of different activities, including task switching. This strategy is pervasive in the field and is not different from attributing aspects of memory to the hippocampus or language production to Broca's area. Beyond the well-known, general problems faced by the localization of psychological functions in particular brain areas (see Anderson, 2014), the search for a concrete task-switching mechanism in the brain has met with mixed results. For instance, a meta-analysis by Agatha Lenartowicz, Donald Kalar, Eliza Congdon and Russell Poldrack (2010) found no unique neural signature for task switching. Neither of these competing results is of course definitive, but we think their disparity is an invitation to find alternative solutions and interpretations for the phenomenon of task switching and, consequently, for intentions.

One of these alternatives is a genuinely ecological approach to task switching: task switching does not happen *in the brain*, but within the organism-environment system. In other words, a proper characterization of task switching cannot be provided by the sole appeal to neural resources. Non-neural resources, like (non-neural) bodily and environmental factors–including affordances and enabling constraints (see Raja & Anderson, 2021)–are necessary to make full sense of this and other psychological phenomena. In the context of such an ecological (or at least distributed) strategy, Dennis Proffit and Sally Linkenauger (2013) have coined the phrase "phenotypic reorganization" to describe the way in which organisms organize themselves for various tasks: walking, running, eating, grasping, throwing, etc. This self-organization creates temporary, reproducible synergies in both the body and brain to achieve the task at hand. Motor synergies in the body and neural synergies in the brain (like TALoNS; see Anderson, 2014), but also synergies that cut across the boundaries of the brain-body-environment system (e.g., Baggs, Raja, & Anderson, 2020; Dotov, Nie, & Chemero, 2010) support the performance of the different tasks cognitive systems face in their daily life (see also Bingham, 1988).

One aspect of this "phenotypic" organization is that it can be re-organized when different conditions are met. A cognitive system organized in this way for a particular task is robust enough to handle different perturbations within the task. For instance, a runner is able to keep running even when several conditions of the ground (e.g., shape, consistency, etc.) change. This is due to the redundancy of bodily and neural resources of the system that, when combined with the constraints of the environment, allow for a stable "running" organization to emerge even if when looking at little details of the system (i.e., some conditions of the ground or the particular activation of a set of muscles) nothing remains equal and constant variability is the rule. The system, though robust, is however not resilient to all possible perturbations and some of them do affect it. When they occur, they can cause the organism to switch tasks, say from grazing to running from a predator or from running to jumping in the face of an obstacle on the ground. It is also, at least in some cases, possible for an organism to resist responding to a perturbation, depending on the intentions of the organism and the particularities of the environmental situation (Withagen, Araújo, & de Poel, 2017). It is likely that, within this general process that cuts across the boundaries of the brain-body-environment system, the insula, for instance, registers and resonates to what changes in the landscape, but this is not the same as saying that insula causes the switch, or *is* the switch, even in cases when it is an important part of the whole causal chain. Thus, in this conceptualization of task switching, no clear distinction can be made between perception, action, or cognitive (e.g., switching) "parts". There is no privileged (e.g., neural) scale (Carello, Vaz, Blau, & Petrusz, 2012), and, it may be hard if not impossible to disassemble the system into stable functional parts that stay within the boundary of brain or skin (de Wit & Matheson, 2022). On the contrary, what relevant perturbations do is to change a given phenotypic organization by disrupting the (enabling) constraints that define its softly-assembled synergetic state, allowing for (or in some cases forcing) a phenotypic reorganization into a new synergy. This can occur in many different systems, some of them non-cognitive. For instance, if you somehow break the toilet bowl during a flush, the constraints imposed by the bowl and drain disappear, and gravity becomes the dominant force, spilling the water onto the bathroom floor. But the process also occurs in the case of cognitive systems executing different tasks. For example, if the blacksmith drops the hot iron he is molding, he ceases to be a "hammerer", and becomes a "hot horseshoe avoider". The carpenter's broken nail will similarly disrupt the hammering while he goes in search of his pliers to remove the shard. As illustrated in these examples, affordance perception and actualization–and switches between them–are typically *dynamic*. Although many affordance judgment paradigms might imply otherwise, affordances are very commonly responded to (through action or inaction) "in the moment".

We have discussed situations in which task switching occurs and can be explained not in terms of an intrinsic mechanism but in terms of the self-organization of the brain-body-environment system, including changes at all scales of that system when it is disrupted. Some of the possible changes are, of course, changes in affordances. And these changes at different scales can be investigated empirically as cognitive systems engage with different tasks and exhibit different intentions.

**How do systems engage with affordances as they move amongst tasks and intentions?**

As argued in the previous Section, a complete answer to the question of how systems switch from responding from one to another affordance requires an analysis of the system at all scales (de Wit & Withagen, 2019). So far, ecological psychology has largely focused on the behavioral scale and hence there is a paucity of research on the contributions of the nervous system in brain-body-environment systems when organisms respond to affordances. However, interest in the neural scale is growing in the ecological community under the header of ecological neuroscience (e.g., Anderson, 2014; de Wit et al., 2017; de Wit & Withagen, 2019; Jirsa, McIntosh, & Huys, 2019; Falandays, Yoshimi, Warren, & Spivey, 2023; Favela, 2024; Raja, 2018, 2020; van der Weel, Agyei, & van der Meer, 2019). Moreover, there is a substantial ecological psychological literature that has focused on characterizing (strategy) switching within various cognitive tasks at the behavioral scale which, although not focused on affordances, is relevant to this question and which we will discuss here (Anastas, Kelty-Stephen, & Dixon, 2011; Bruineberg, Seifert, Rietveld, & Kiverstein, 2021; Dixon, Kelty-Stephen, & Anastas, 2014; Dotov et al., 2010; Favela, Amon, Lobo, & Chemero, 2021; Kello & Van Orden, 2009; Nalepka et al., 2017, 2019, 2021; Patil, Nalepka, Kallen, & Richardson, 2020; Stephen, Boncoddo, Magnuson, & Dixon, 2009; Stephen, Dixon, & Isenhower, 2009; Van Orden, Kloos, & Wallot, 2011; Wallot, Kee, & Kelty-Stephen, 2019). We will end the chapter by discussing ongoing research in the lab of one of us that characterizes the neural scale while participants dynamically respond to affordances.

Anastas et al. (2011) presented participants with a modified version of the Wisconsin card sorting test, which is traditionally used to test childrens' ability to flexibly switch amongst sorting rules. In this modified version, young adults were instructed to induce the correct sorting rule on the basis of feedback about their card placement performance. In the language introduced above, in the early phase of this task, as participants were attempting to discover the rule, their phenotypic organization might be described as a "rule-searching card sorter". Then, following induction of the correct sorting rule, they can be seen as having reorganized into a "rule-following cart sorter". The question is how to understand this switch. Is it best conceptualized intrinsically, as a switch controlled by activity at the neural scale (cf. Menon & Uddin, 2010), or as a switch between two self-organizing brain and body-spanning synergies? Results suggest the latter. As participants grasped and placed cards, their hand movements were tracked at a 60 Hz sampling rate and with millimeter precision. These time series were then submitted to detrended fluctuation analysis (DFA; Peng et al., 1994). DFA is used to estimate Hurst exponents, which characterize the degree to which the dynamics of fluctuations within time series are scale-free or "fractal", meaning that the statistics of fluctuations at any scale resemble those of fluctuations at all other available scales. This is indicative of interactions across scales and, with that, the absence of a privileged scale. Furthermore, an increase in Hurst exponent towards a value of 1 indicates loosening of constraints within a system (such as a "card sorter"), allowing for reorganization of that system, while a decrease in the direction of 0 indicates increased stability in the system (Hardstone et al., 2012) and thus presumably less openness to perturbation. Results showed that the Hurst exponent derived, recall, from participants' hand motions showed a sharp rise and then fall as participants induced the rule, suggesting, first, that they transitioned from one stable phenotypic organization into another and, second, interactions across scales. These results simultaneously argue against both an intrinsic view of switching (in this case between strategies) and a clean localization of action and cognitive processes onto specialized parts within scales (Dixon et al., 2014).

Thus, there is evidence for phenotypic organization and reorganization in the non-neural body (e.g., Anastas et al., 2011; Stephen et al., 2009). Similar results have been obtained when looking at reorganization of synergies involving not only the body but also aspects of the physical environment (i.e., tools; Dotov et al., 2010; Favela et al., 2021) and the social environment (Nalepka et al., 2017, 2019, 2021; Patil, et al., 2020). There is also a large literature looking at flexible reorganization in the brain (e.g., Cocchi, Gollo, Zalesky, & Breakspear, 2017; Kelso, 2012). Though, taken together, all of the above implies that reorganization involves soft-assembly of synergies that span the neural and behavioral scales, very little research has attempted to characterize the fractality of time series simultaneously at each of these different scales (cf. Kello & Van Orden, 2009). Furthermore, to our knowledge no one has investigated such organization and reorganization while participants dynamically respond to affordances.

We are currently developing a project that attempts to do just this. In an augmented reality interception task, participants are presented with expanding dot patterns that create the impression of a looming ball superimposed on a large response button attached to the wall in front of them. Balls approach the participant at a wide range of randomized speeds and the participant attempts to intercept the ball by hitting the button at the perceived time of impact of the ball with the wall. In an initial training phase, participants are instructed to always attempt to intercept the ball. Successful interceptions are defined as hitting the response button within a window of 140 ms around time-to-contact (TTC) of the ball relative to the wall of zero. Participants receive performance feedback allowing them to discover which balls they can and cannot intercept successfully, i.e., their affordance boundaries. In a subsequent affordance perception and actualization phase in which they no longer receive feedback, participants are instructed to "only go for balls that you can successfully intercept". This paradigm allows us to track participants as they switch from being capable to incapable "interceptors", and the behavioral and neural characteristics that are associated with sensitivity to this boundary. We are using functional near-infrared spectroscopy (fNIRS) to characterize the neural scale during the affordance perception and actualization phase. fNIRS allows for movement in participants (Pinti et al., 2020) making it a suitable method for measuring neural activity in dynamic affordance tasks. Our montage is set up to capture signal from "sensorimotor" cortex extending into the frontal cortex, given arguments for the involvement of those regions in affordance perception and actualization (e.g., Pezullo & Cisek, 2016).

We predict an increase in the Hurst exponent of the fNIRS time series as participants approach their affordance boundaries, since this is where we would expect changes in the stability of the brain-body-environment system as it reorganizes. Changes in information use (i.e., towards TTC-specifying information) as well as changes in interception skill and affordance boundary sensitivity can also be conceptualized as phenotypic reorganizations and so we also predict changes in Hurst exponents as a function of those (cf. Hajnal, Clark, Doyon, & Kelty-Stephen, 2018; Mangalam, Chen, McHugh, Singh, & Kelty-Stephen, 2020; Stephen, Arzamarski, & Michaels, 2010; Palatinus, Dixon, & Kelty-Stephen, 2013). Tests of some of these predictions are currently underway. Future work will characterize scale-free fluctuations simultaneously at the behavioral and neural scales by including analyses of arm, head and eye movements in addition to fNIRS time series. Such a finding would support the conceptualization of brains–not as intrinsic controllers of switches–but as components of larger self-organizing brain and body spanning soft-assembled synergies that reorganize as a function of changes in the affordances available to them, as outlined in this chapter.

Finally, we briefly note that we have not discussed a deeper question underlying the question of why we expect scale-free dynamics at the neural scale: *why are brains present at all* in many organisms, given that transitions between different behavioral states are also possible in systems without nervous systems (Fultot, Frazier, Turvey, & Carello, 2019). One start of an answer to this question might be that neurons and neural networks are an excellent substrate for the complex, dynamic interactions between elements of a system that are required for flexible self-organization and reorganization (Cocchi et al., Kelso, 2012); another, that dynamics at the neural scale enable functional organization and reorganization at very short time scales (Fultot et al., 2019) which arguably constitutes an adaptive advantage to many forms of life.

**Acknowledgements**

We thank Michael Anderson for fruitful discussions and for his suggestion to explore task switching in the context of affordance theory and research. V.R.'s research is funded by the Juan de la Cierva-Incorporación Research Program and the project PID2021-127294NA-I00 of the Spanish Ministry of Science.